\documentclass[prb,twocolumn,amsmath,amssymb,superscriptaddress, bibnotes]{revtex4-2}
\usepackage{graphicx}
\usepackage{dcolumn}
\usepackage{bm}
\usepackage{color}
\usepackage{braket} 
\usepackage{booktabs}

\begin{document}

\title{Role of $d$-electron density of states in the quantum size effect \newline of Pt-Ni and Pt-Pd nanoparticles}

\author{Shunsaku~Kitagawa}
\email{kitagawa.shunsaku.8u@kyoto-u.ac.jp}

\author{Taishi~Ihara}
\author{Yudai~Kinoshita}
\author{Kenji~Ishida}

\affiliation{Department of Physics, Kyoto University, Kyoto 606-8502, Japan}

\author{Kouhei~Kusada}

\affiliation{Department of Chemistry, Kyoto University, Kyoto 606-8502, Japan}
\affiliation{The Hakubi Center for Advanced Research, Kyoto University, Kyoto 606-8501, Japan}
\affiliation{Institute for Advanced Study, Kyushu University, Fukuoka 816-8580, Japan}
\affiliation{Faculty of Engineering Sciences, Kyushu University, Fukuoka 816-8580, Japan}

\author{Hiroshi~Kitagawa}

\affiliation{Department of Chemistry, Kyoto University, Kyoto 606-8502, Japan}

\date{\today}

\begin{abstract}
We investigated the quantum size effect (QSE) in bimetallic Pt$_{1-x}$Pd$_x$ and Pt$_{1-x}$Ni$_x$ nanoparticles, using $^{195}$Pt nuclear magnetic resonance measurements.
The temperature and size dependencies of the anomaly in the nuclear spin-lattice relaxation rate divided by temperature $1/T_1T$ in the Pt$_{1-x}$Pd$_x$ nanoparticles suggest similar electron states between Pt and Pd atoms and are well understood by the QSE.
The temperature and composition variations of $1/T_1T$ and Knight shift reveal a systematic increase in the density of states and reduction of the characteristic energy scale $T^*$ with increasing Ni content, consistent with the Kubo gap $\delta_{\mathrm{Kubo}}$.
In contrast to Pt$_{1-x}$Cu$_x$ nanoparticles where the QSE is suppressed, the Pt$_{1-x}$Ni$_x$ nanoparticles exhibit clear signatures of quantum energy discretization.
This discrepancy highlights the essential role of $d$-electrons in the manifestation of the QSE.
Furthermore, analysis of the modified Korringa parameter $K(\alpha)$ suggests enhanced ferromagnetic correlations with increasing Ni concentration, approaching a ferromagnetic quantum critical regime.
These results provide experimental evidence that $d$-electron density of states plays a crucial role in the manifestation of the QSE in the nanoparticles formed by the metallic $d$-electron atoms.
\end{abstract}

\maketitle

\section{Introduction}
Nanoparticles, known as quantum dots, are particles with diameters on the order of nanometers~\cite{C.B.Murray_ARMS_2000}.
The nanoparticles exhibit properties distinct from their bulk counterparts, owing to a large surface-to-volume ratio~\cite{E.Roduner_CSR_2006} and the emergence of quantum size effects (QSE)~\cite{A.I.Ekimov_JETPL_1981,Rossetti1983,J.A.A.J.Perenboom_PhysRep_1981}.
Thus, they have been studied~\cite{R.Kubo_JPSJ_1962,L.P.Gorkov_JETP_1965,W.P.Halperin_RMP_1986} and utilized in a wide range of fields~\cite{A.P.Alivisatos_science_1996,Seh2017,V.Harish_nano_2022,Antoine2023}.
In metallic systems, the QSE arises from the discretization of electronic states, as first predicted by Kubo~\cite{R.Kubo_JPSJ_1962}.
According to this theory, the mean energy level spacing $\delta_{\mathrm{Kubo}}$ is inversely proportional to the product of the number of atoms $N$ in nanoparticle and the density of states at the Fermi energy $D(E_{\mathrm{F}})$ of the metallic atoms, given by $\delta_{\mathrm{Kubo}} = [ND(E_{\mathrm{F}})]^{-1}$.

Experimental verification of the QSE in metallic nanoparticles~\cite{Y.Volokitin_Nature_1996} has been complicated by the surface effects, which often obscure the intrinsic quantum confinement.
Surface oxidation and other chemical degradation can alter the metallic nature at the surface, introducing ambiguity in the interpretation of experimental data~\cite{Bucher1989,T.Fujii_PRB_2022}.

$^{195}$Pt-NMR studies on Pt nanoparticles have successfully separated the properties arising from surface and interior regions, due to the high $D(E_{\mathrm{F}})$ of the Pt atoms~\cite{H.E.Rhodes_PRB_1982,T.Okuno_PRB_2020}.
Anomalous increases in nuclear spin-lattice relaxation rate divided by temperature $1/T_1T$ below $T^*$, observed for both surface and interior signals, strongly support the presence of the Kubo gap $\delta_{\mathrm{Kubo}}$.
Systematic particle-size and magnetic-field dependencies of $T^*$ further confirm its correlation with $\delta_{\mathrm{Kubo}}$~\cite{T.Okuno_PRB_2020,T.Okuno_JPSJ_2020}.
Moreover, Fujii $et~al.$ also reported the emergence of a Kubo gap in thiol-capped Pt nanoparticles~\cite{T.Fujii_PRB_2022}.

To understand the effect of electronic structure on the QSE, the Pt$_{1-x}$Cu$_x$ nanoparticles, in which Cu primarily contributes $s$-electron states, were studied~\cite{B.Hammer_Nature_1995}.
In the Pt$_{1-x}$Cu$_x$ nanoparticles, no significant $T^*$ variation against $x$ was observed, but the signatures of the QSE were diminished at high Cu concentrations~\cite{S.Kitagawa_PRB_2024}.
These findings imply that the importance of $d$-electrons, such as in the Pt atoms, is crucial for the emergence of the QSE.

\begin{figure}[!tb]
\includegraphics[width=8.5cm,clip]{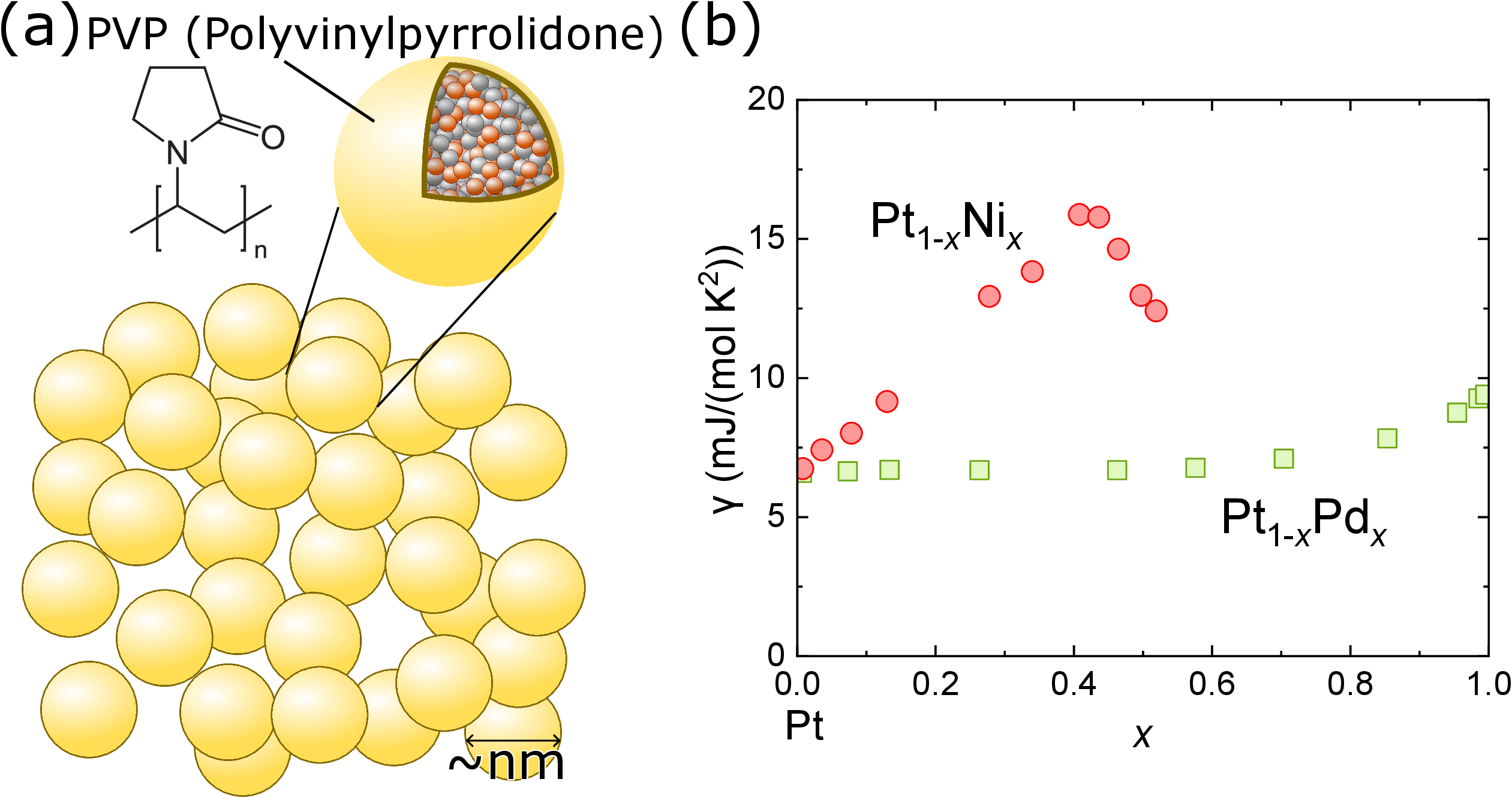}
\caption{
(a) Schematic image of Pt$_{1-x}$$X$$_{x}$ nanoparticles ($X =$ Pd, and Ni).
Nanoparticles are covered by polyvinylpyrrolidone to prevent oxidation and particle-to-particle contact. 
(b) $x$ dependence of the estimated Sommerfeld coefficient, which is proportional to the density of states at Fermi energy $D(E_{\rm F})$ in bulk Pt$_{1-x}$$X$$_{x}$ alloy~\cite{J.Inoue_JPSJ_1977}.
}
\label{Fig.1}
\end{figure}

\begin{table*}[htb]
    \centering
    \caption{Synthesis conditions of the Pt$_{1-x}$Pd$_x$ and
    Pt$_{1-x}$Ni$_x$ nanoparticles.
    Particle diameters are given as the mean $\pm$ standard deviation
    determined from the TEM images.
    For Pt$_{0.53}$Pd$_{0.47}$, Pt$_{0.80}$Pd$_{0.20}$, and
    Pt$_{0.68}$Ni$_{0.32}$, the synthesis was repeated three times,
    and the resulting three batches were collected and combined into
    one sample for each composition.}
    \label{table}

    \resizebox{\textwidth}{!}{
    \begin{tabular}{l c c l c l l l}
        \toprule
        Sample
        & \begin{tabular}[c]{@{}c@{}}Mean diameter\\(nm)\end{tabular}
        & \begin{tabular}[c]{@{}c@{}}Temp.\\($^\circ$C)\end{tabular}
        & \begin{tabular}[c]{@{}l@{}}Reductant\\(mL)\end{tabular}
        & \begin{tabular}[c]{@{}c@{}}PVP\\(mmol)\end{tabular}
        & \begin{tabular}[c]{@{}l@{}}Pt precursor\\(mmol)\end{tabular}
        & \begin{tabular}[c]{@{}l@{}}Pd or Ni precursor\\(mmol)\end{tabular}
        & \begin{tabular}[c]{@{}l@{}}Precursor solvent\\(mL)\end{tabular} \\
        \midrule

        $\mathrm{Pt}_{0.44}\mathrm{Pd}_{0.56}$
        & $5.4 \pm 0.7$
        & 175
        & \begin{tabular}[c]{@{}l@{}}
          Ethylene glycol (Wako)\\200
          \end{tabular}
        & 10.0
        & \begin{tabular}[c]{@{}l@{}}
          $\mathrm{K}_2[\mathrm{PtCl}_4]$ (Wako)\\1.00
          \end{tabular}
        & \begin{tabular}[c]{@{}l@{}}
          $\mathrm{K}_2[\mathrm{PdCl}_4]$ (Wako)\\1.00
          \end{tabular}
        & \begin{tabular}[c]{@{}l@{}}
          $\mathrm{H}_2\mathrm{O}$\\50
          \end{tabular} \\

        \addlinespace
        $\mathrm{Pt}_{0.53}\mathrm{Pd}_{0.47}$
        & $4.3 \pm 0.7$
        & 175
        & \begin{tabular}[c]{@{}l@{}}
          Ethylene glycol (Wako)\\200
          \end{tabular}
        & 4.0
        & \begin{tabular}[c]{@{}l@{}}
          $\mathrm{K}_2[\mathrm{PtCl}_4]$ (Wako)\\0.40
          \end{tabular}
        & \begin{tabular}[c]{@{}l@{}}
          $\mathrm{K}_2[\mathrm{PdCl}_4]$ (Wako)\\0.40
          \end{tabular}
        & \begin{tabular}[c]{@{}l@{}}
          $\mathrm{H}_2\mathrm{O}$\\60
          \end{tabular} \\

        \addlinespace
        $\mathrm{Pt}_{0.80}\mathrm{Pd}_{0.20}$
        & $4.6 \pm 0.5$
        & 175
        & \begin{tabular}[c]{@{}l@{}}
          Ethylene glycol (Wako)\\200
          \end{tabular}
        & 4.0
        & \begin{tabular}[c]{@{}l@{}}
          $\mathrm{K}_2[\mathrm{PtCl}_4]$ (Wako)\\0.64
          \end{tabular}
        & \begin{tabular}[c]{@{}l@{}}
          $\mathrm{K}_2[\mathrm{PdCl}_4]$ (Wako)\\0.16
          \end{tabular}
        & \begin{tabular}[c]{@{}l@{}}
          $\mathrm{H}_2\mathrm{O}$\\60
          \end{tabular} \\

        \addlinespace
        $\mathrm{Pt}_{0.68}\mathrm{Ni}_{0.32}$
        & $3.8 \pm 0.4$
        & 240
        & \begin{tabular}[c]{@{}l@{}}
          Triethylene glycol (Wako), 300\\
          + NaOH (Wako), 0.6 mmol
          \end{tabular}
        & 10.0
        & \begin{tabular}[c]{@{}l@{}}
          $\mathrm{Pt(acac)}_2$ (Wako)\\0.23
          \end{tabular}
        & \begin{tabular}[c]{@{}l@{}}
          $\mathrm{Ni(acac)}_2$ (Aldrich)\\0.10
          \end{tabular}
        & \begin{tabular}[c]{@{}l@{}}
          Tetraethylene glycol (Wako)\\6
          \end{tabular} \\

        \addlinespace
        $\mathrm{Pt}_{0.71}\mathrm{Ni}_{0.29}$
        & $9.9 \pm 3.4$
        & 270
        & \begin{tabular}[c]{@{}l@{}}
          Tetraethylene glycol (Wako)\\150
          \end{tabular}
        & 5.0
        & \begin{tabular}[c]{@{}l@{}}
          $\mathrm{Pt(acac)}_2$ (Wako)\\0.45
          \end{tabular}
        & \begin{tabular}[c]{@{}l@{}}
          $\mathrm{Ni(acac)}_2$ (Aldrich)\\0.19
          \end{tabular}
        & \begin{tabular}[c]{@{}l@{}}
          Tetraethylene glycol (Wako)\\6
          \end{tabular} \\

        \addlinespace
        $\mathrm{Pt}_{0.95}\mathrm{Ni}_{0.05}$
        & $7.3 \pm 2.4$
        & 270
        & \begin{tabular}[c]{@{}l@{}}
          Tetraethylene glycol (Wako)\\150
          \end{tabular}
        & 5.0
        & \begin{tabular}[c]{@{}l@{}}
          $\mathrm{Pt(acac)}_2$ (Wako)\\0.50
          \end{tabular}
        & \begin{tabular}[c]{@{}l@{}}
          $\mathrm{Ni(acac)}_2$ (Aldrich)\\0.03
          \end{tabular}
        & \begin{tabular}[c]{@{}l@{}}
          Tetraethylene glycol (Wako)\\6
          \end{tabular} \\

        \bottomrule
    \end{tabular}
    }
\end{table*}

\begin{figure*}[!tb]
\includegraphics[width=16cm,clip]{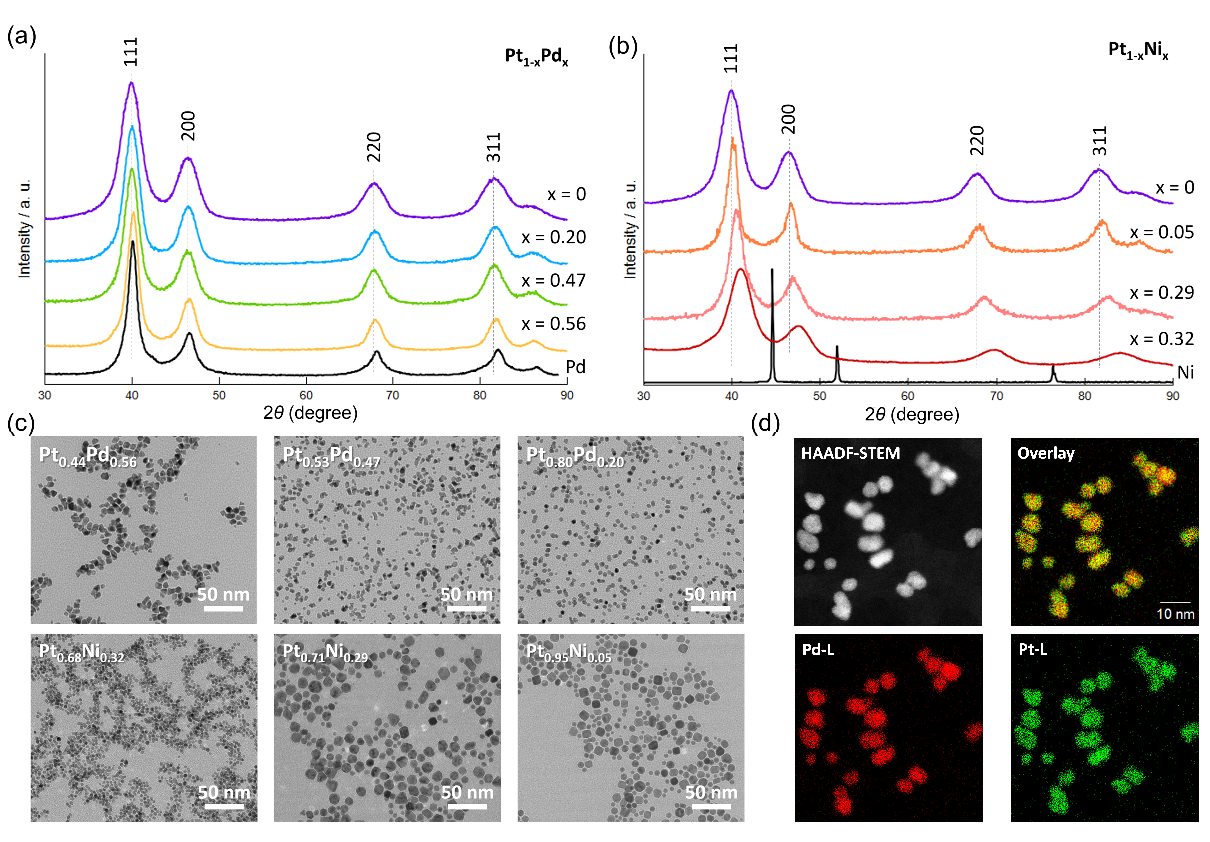}
\caption{
Powder X-ray diffraction patterns of (a) Pt$_{1-x}$Pd$_{x}$ and (b) Pt$_{1-x}$Ni$_{x}$ using Cu-K$\alpha$ radiation. 
(c) TEM images of the synthesized nanoparticles.
(d) High-angle annular dark field scanning TEM image of Pt$_{0.53}$Pd$_{0.47}$, and the corresponding energy dispersive X-ray maps of Pd-L and Pt-L. The overlay of the Pd-L and Pt-L maps shows the formation of the solid-solution alloy.
}
\label{Fig.1-2}
\end{figure*}

In this context, we explore the influence of the $d$-electron hybridization on the QSE by investigating Pt$_{1-x}$Pd$_x$ and Pt$_{1-x}$Ni$_x$ nanoparticles [Fig.~\ref{Fig.1}(a)].
In the bulk systems, Ni atoms have a larger density of states at the Fermi level than Pt, and are known to exhibit strong ferromagnetic correlations~\cite{B.XavierBatlle_JPhysD_2002}, while $D(E_{\mathrm{F}})$ of Pd is almost similar to that of Pt~\cite{J.Inoue_JPSJ_1977}, as shown in Fig.~\ref{Fig.1}(b).
Our systematic NMR results on the Pt$_{1-x}$Pd$_x$ and Pt$_{1-x}$Ni$_x$ nanoparticles reveal that the Ni substitution systematically reduces $T^*$ and enhances the Knight shift, consistent with an increased $D(E_{\mathrm{F}})$ and Kubo's prediction, although the results in the Pt$_{1-x}$Pd$_x$ nanoparticles are well understood by the QSE without changing in $D(E_{\mathrm{F}})$. 
In contrast to the Pt$_{1-x}$Cu$_x$ nanoparticles~\cite{S.Kitagawa_PRB_2024}, the Pt$_{1-x}$Ni$_x$ nanoparticles show that the QSE are related to $D(E_{\mathrm{F}})$ of the mixing atoms, confirming the importance of $d$-electron involvement.

\section{Experimental Methods}
In a typical synthesis of the Pt$_{1-x}$Pd$_x$ and Pt$_{1-x}$Ni$_x$ nanoparticles, poly(N-vinyl-2-pyrrolidone) (PVP K-30, Wako) was dissolved in a reductant. 
Metal precursors were dissolved in deionized water or tetraethylene glycol. 
The aqueous precursor solution was then slowly added to the heated reductant solution at a set temperature with magnetic stirring. 
The solution was maintained at the set temperature while adding the solution. 
After adding the precursor solution, the mixture was kept stirring for 10 min and then cooled to room temperature. 
The synthesized nanoparticles were separated by centrifuging. 
The details of the amount of chemicals are summarized in Table~\ref{table}. 
X-ray diffraction (XRD) analysis revealed that all the samples have a face-centered cubic crystal structure, and the peak shift to a higher angle with increasing Pd or Ni composition indicated the formation of solid-solution alloys [Figs.~\ref{Fig.1-2}(a) and ~\ref{Fig.1-2}(b)]. 
The particle diameters were evaluated from the TEM images [Fig.~\ref{Fig.1-2}(c)], and the mean diameters and standard deviations are summarized in Table~\ref{table}.
The metal composition was determined by X-ray fluorescence (XRF) analysis using several tens of milligrams of sample and an X-ray beam approximately 10~mm in diameter.
Thus, the composition determined by XRF represents an ensemble average over a large number of nanoparticles.
Figure \ref{Fig.1-2}(d) shows elemental maps of Pt$_{0.53}$Pd$_{0.47}$ as an example confirming the formation of a solid-solution alloy.
Quantitative energy-dispersive X-ray spectroscopy analyses of individual nanoparticles revealed particle-to-particle compositional variations of several atomic percent around the XRF-determined average composition.

$^{195}$Pt ($I = 1/2$, $^{195}\gamma/2\pi = 9.153$~MHz/T) NMR measurements were carried out using a conventional spin-echo technique~\cite{R.K.Harris_2001,N.J.Stone_Q_2016}.
The magnetic field-swept NMR spectra were obtained by recording the spin-echo signal observed after a standard $\pi/2$--$\pi$ radio frequency pulse sequence at 111~MHz for $^{195}$Pt-NMR.
The NMR spectra are presented with the horizontal axis representing $K = (f - f_0)/f_0$.
Here, $f_0 = (\gamma/2\pi)\mu_0H$.
The magnetic field was calibrated using a $^{63(65)}$Cu [$^{63(65)}\gamma/2\pi = 11.285 (12.089)$~MHz/T]-NMR signal with the Knight shift $K_{\rm Cu} = 0.238$\% from a NMR coil~\cite{Metallicshifts_1977}.
$T_1$ was measured using the saturation-recovery method.
At high temperatures, single exponential recovery was observed, while at low temperatures, multicomponent relaxation was fitted with the stretched exponential fitting:
\begin{align}
    M(t) = M(\infty)\left[ 1 - \exp\left\{ -\left( \frac{t}{T_1} \right)^\beta \right\} \right],
\end{align}
where $\beta = 1$ corresponds to a single component and $\beta < 1$ indicates relaxation inhomogeneity arising from the multicomponents of $T_1$.

\begin{figure}[!tb]
\includegraphics[width=7.5cm,clip]{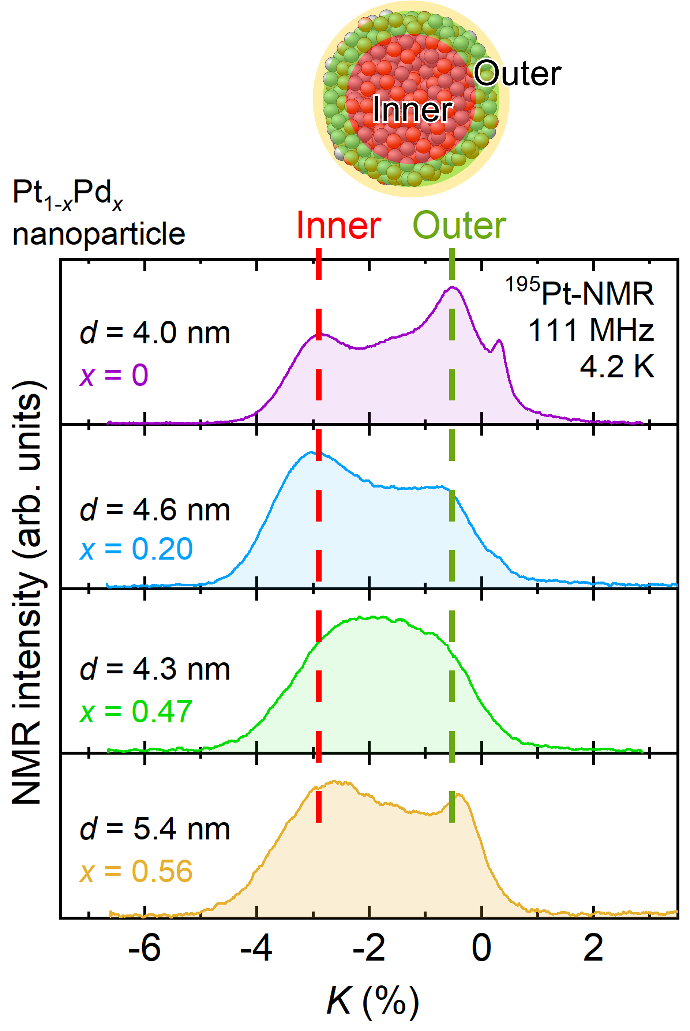}
\caption{
$^{195}$Pt-NMR spectrum of Pt$_{1-x}$Pd$_{x}$ nanoparticles for various $x$ measured at 4.2~K.
The particle size in each $x$ is indicated in the panel.
For $^{195}$Pt-NMR, the two-peak structure originating from the inner and the outer atoms was observed, indicated by the dashed lines.
For reference, the data for the monometallic Pt nanoparticles~\cite{T.Okuno_PRB_2020} are also presented.
}
\label{Fig.2}
\end{figure}
\section{Results}
\subsection{Pt$_{1-x}$Pd$_x$ Nanoparticles}

Figure~\ref{Fig.2} shows the $^{195}$Pt-NMR spectra at 4.2~K for Pt$_{1-x}$Pd$_x$ nanoparticles with various Pd concentrations.
The spectrum of monometallic Pt nanoparticles with a 4.0~nm diameter is also shown for comparison.
Each spectrum consists of two distinguishable peaks, attributed to the outer and inner regions of the nanoparticles, respectively.
The Knight-shift difference between the peaks reflects the local variation in the electronic density of states $D(E_{\mathrm{F}})$.
Compared with the 4~nm Pt nanoparticles, the relative intensity of the outer peak is smaller in Pt$_{1-x}$Pd$_x$ samples, suggesting a larger particle size, consistent with the results of transmission electron microscopy.

\begin{figure}[!tb]
\includegraphics[width=\linewidth,clip]{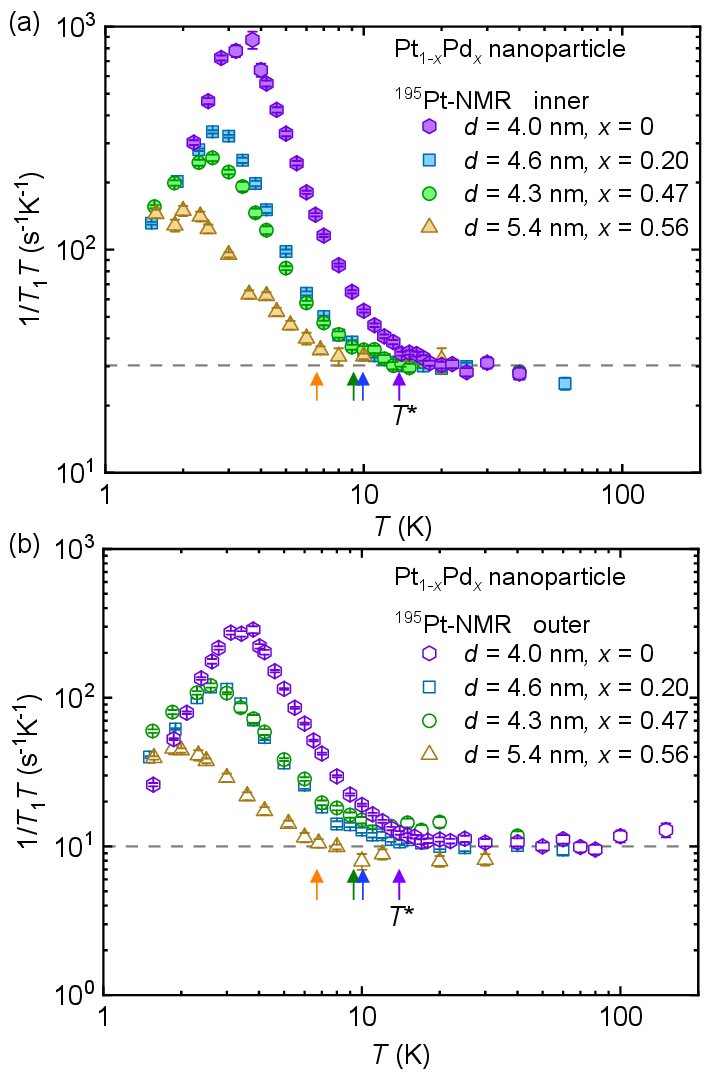}
\caption{
Temperature dependence of $1/T_1T$ in the Pt$_{1-x}$Pd$_x$ nanoparticles at various $x$ for the (a) inner and (b) outer regions of nanoparticles.
The dashed lines indicate a constant value at high temperatures.
The arrows indicate $T^*$, where $1/T_1T$ starts to increase.
For reference, the data for the monometallic Pt nanoparticles~\cite{T.Okuno_PRB_2020} are also presented.
}
\label{Fig.3}
\end{figure}

Figures~\ref{Fig.3}(a) and \ref{Fig.3}(b) show the temperature dependence of the nuclear spin-lattice relaxation rate divided by temperature $1/T_1T$ measured at the peaks arising from the inner and outer region, respectively.
In all Pt$_{1-x}$Pd$_x$ samples, $1/T_1T$ is constant at high temperatures, indicating metallic behavior.
On cooling, a pronounced increase in $1/T_1T$ was observed below a characteristic temperature $T^*$, followed by a maximum.
This behavior is consistent with the existence of Kubo gap $\delta_{\mathrm{Kubo}} \sim k_{\mathrm{B}}T^*$ observed in the Pt nanoparticles\cite{T.Okuno_PRB_2020}.

Despite the compositional changes, the temperature dependence of $1/T_1T$ is qualitatively similar between the inner and outer regions, reinforcing the attribution of the low-temperature anomaly to the QSE rather than surface-related effects.
The both reductions in $T^*$ and the peak in $1/T_1T$ for Pt$_{1-x}$Pd$_x$ samples, compared to Pt nanoparticles with particle size $d$ = 4.0 nm, are ascribed to the larger particle size and not to changes in $D(E_{\mathrm{F}})$, as Pt and Pd have a similar electronic structures.
The smaller magnitude of the $1/T_1T$ peak with Pd concentration is due to particle size, as observed in the size-dependence of $1/T_1T$ for the Pt nanoparticles\cite{T.Okuno_PRB_2020}. 

\begin{figure}[!tb]
\includegraphics[width=\linewidth,clip]{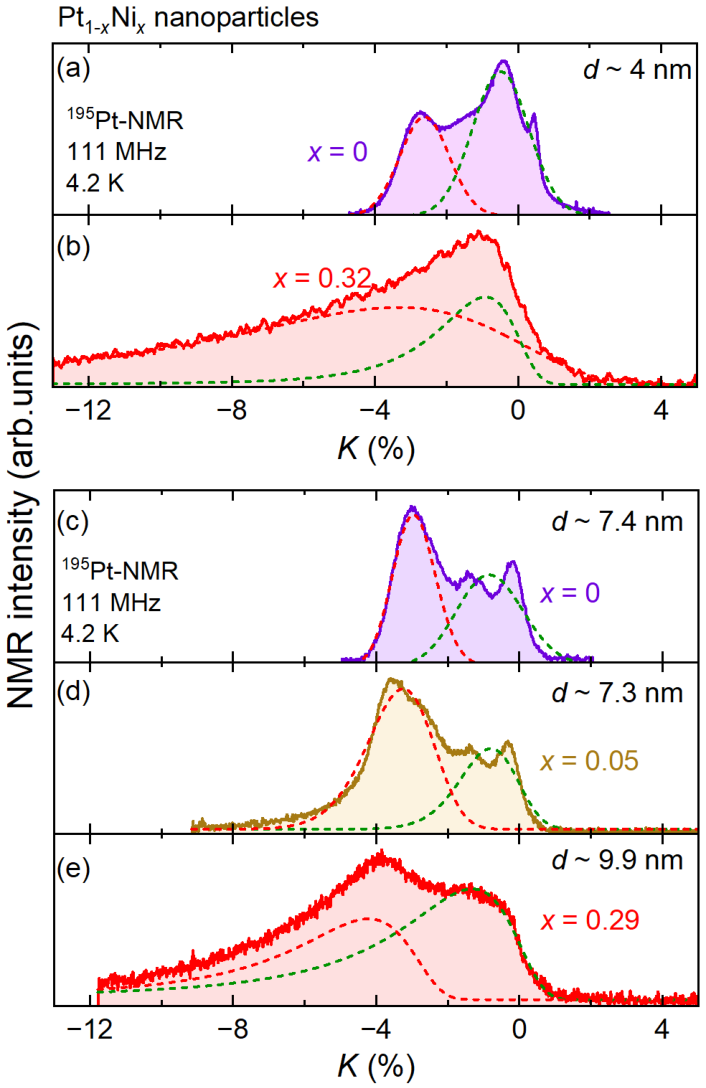}
\caption{
$^{195}$Pt-NMR spectrum of Pt$_{1-x}$Ni$_{x}$ nanoparticles for various $x$ measured at 4.2~K.
The particle size in each $x$ is indicated in the panel.
The dominant inner- and outer-region components are fitted by asymmetric double-peak functions, as indicated by the dashed curves. Minor oxide and intermediate-layer components discussed in the text are not included in the fits.}
For reference, the data for the monometallic Pt nanoparticles~\cite{T.Okuno_PRB_2020} are also presented.
\label{Fig.4}
\end{figure}

\subsection{Pt$_{1-x}$Ni$_x$ Nanoparticles}

Figure~\ref{Fig.4} displays the $^{195}$Pt-NMR spectra at 4.2~K for Pt$_{1-x}$Ni$_x$ nanoparticles with $d = 4$~nm
[Figs.~\ref{Fig.4}(a) and \ref{Fig.4}(b)] and 7-10~nm [Figs.~\ref{Fig.4}(c)--\ref{Fig.4}(e)].
With increasing Ni content, the spectra develop a pronounced tail toward negative Knight shifts compared with the Pt nanoparticles.
This asymmetric broadening reflects an increase in the local density of states near the Fermi energy induced by Ni substitution.
To estimate representative peak positions for the inner and outer regions, each spectrum was fitted by the sum of two asymmetric Gaussian functions:
\begin{equation}
 I(K) = I_0 + \sum_{i=1}^{2} \frac{A_i}{\sqrt{2\pi}\omega_i} \exp\left[ -\frac{1}{2} \left\{ \frac{K-K_i} {\omega_i+\alpha_i(K-K_i)} \right\}^{2} \right],
 \label{eq:asymmetric_gaussian}
\end{equation}
where $I_0$ is a constant background, and $A_i$, $K_i$, $\omega_i$, and $\alpha_i$ denote the intensity, peak position, characteristic width, and asymmetry parameter of the $i$th component, respectively.
The two components, shown by the dashed curves in Fig.~\ref{Fig.4}, are assigned to the inner and outer regions of the nanoparticles.
In particular, for $x \simeq 0.3$, the broadening becomes so significant that the two-peak structure observed in the Pt and Pt$_{1-x}$Pd$_x$nanoparticles becomes blurred.
The inner peak is not clearly resolved, especially in the 4-nm nanoparticles.

Small additional features are visible in some of the spectra in Fig.~\ref{Fig.4}.
The weak sharp signal on the positive-shift side of Fig.~\ref{Fig.4}(a), near $K \sim +0.5\%$, is attributed to a small amount of oxidized Pt, most likely H$_2$Pt(OH)$_6$, as reported previously\cite{T.Okuno_PRB_2020}, and is therefore excluded from the fitting.
The intermediate features between the dominant inner and outer peaks in Figs.~\ref{Fig.4}(c) and \ref{Fig.4}(d) are likely associated with Pt nuclei in subsurface or intermediate layers.
The present two-component analysis is intended to estimate representative peak positions for the bulk-like inner and outer regions rather than to provide an exhaustive shell-by-shell decomposition.
Because the intermediate features are weak and strongly overlap with the dominant components, including them does not yield stable and unique fitting parameters within the experimental resolution.
They were therefore omitted from the present analysis.

\begin{figure}[!tb]
\includegraphics[width=\linewidth,clip]{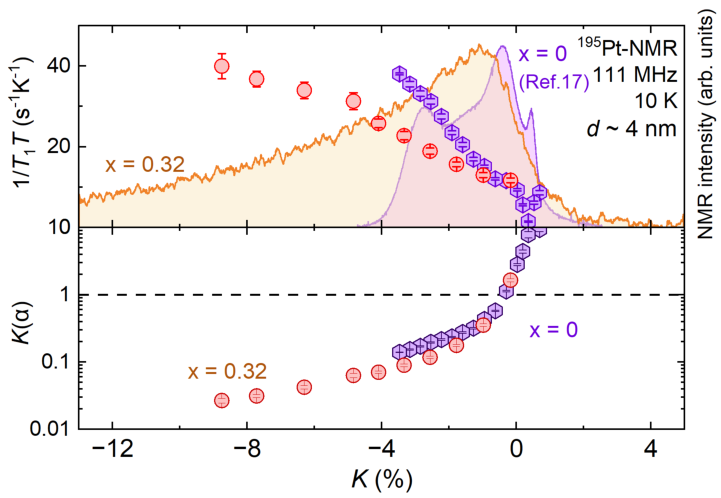}
\caption{
(Top)Position dependence of $1/T_1T$ together with $^{195}$Pt-NMR spectrum for Pt and Pt$_{1-x}$Ni$_{x}$ nanoparticles with $d \sim 4$~nm  measured at 10~K.
(Bottom) Position dependence of $K(\alpha)$ deduced by eq.\eqref{eq.Ka}.
For reference, the data for the monometallic Pt nanoparticles~\cite{T.Okuno_PRB_2020} are also presented.
}
\label{Fig.4-2}
\end{figure}

To evaluate the magnetic correlation strength, we analyzed the site-dependent $1/T_1T$ and Knight shift $K$, and estimated the modified Korringa parameter $K(\alpha)$.
The results are shown in Fig.~\ref{Fig.4-2}.
$K(\alpha)$ is defined as
\begin{align}
    K(\alpha) = \frac{S}{T_1TK_s^2},
    \label{eq.Ka}
\end{align}
where $S$ is the Korringa constant, and $K_s$ is the spin part of the Knight shift.
The Knight shift $K$ is separated into spin and orbital parts: $K = K_{\text{s}} + K_{\text{orb}}$.
Here, $K_{\text{orb}}$ is taken to be +0.46\%\cite{T.Okuno_PRB_2020}.
A value of $K(\alpha) < 1$ indicates ferromagnetic correlations, while $K(\alpha) > 1$ implies antiferromagnetic correlations\cite{T.Moriya_JPSJ_1963}.

In the Pt$_{1-x}$Ni$_x$ nanoparticles, $K(\alpha)$ decreases with increasing $|K|$.
At the high-shift edge ($K \sim -9\%$), $K(\alpha)$ reaches values as low as 0.03, suggesting extremely strong ferromagnetic correlations.
Small $K(\alpha)$ was observed in the Pt nanoparticles, but the Ni substitution makes $K(\alpha)$ more significantly smaller.
In the bulk system, ferromagnetism emerges at $x \sim 0.4$ in Pt$_{1-x}$Ni$_x$~\cite{J.Inoue_JPSJ_1977}, implying that $x = 0.3$ nanoparticles are located near the quantum critical ferromagnetic concentration.

\begin{figure}[!tb]
\includegraphics[width=\linewidth,clip]{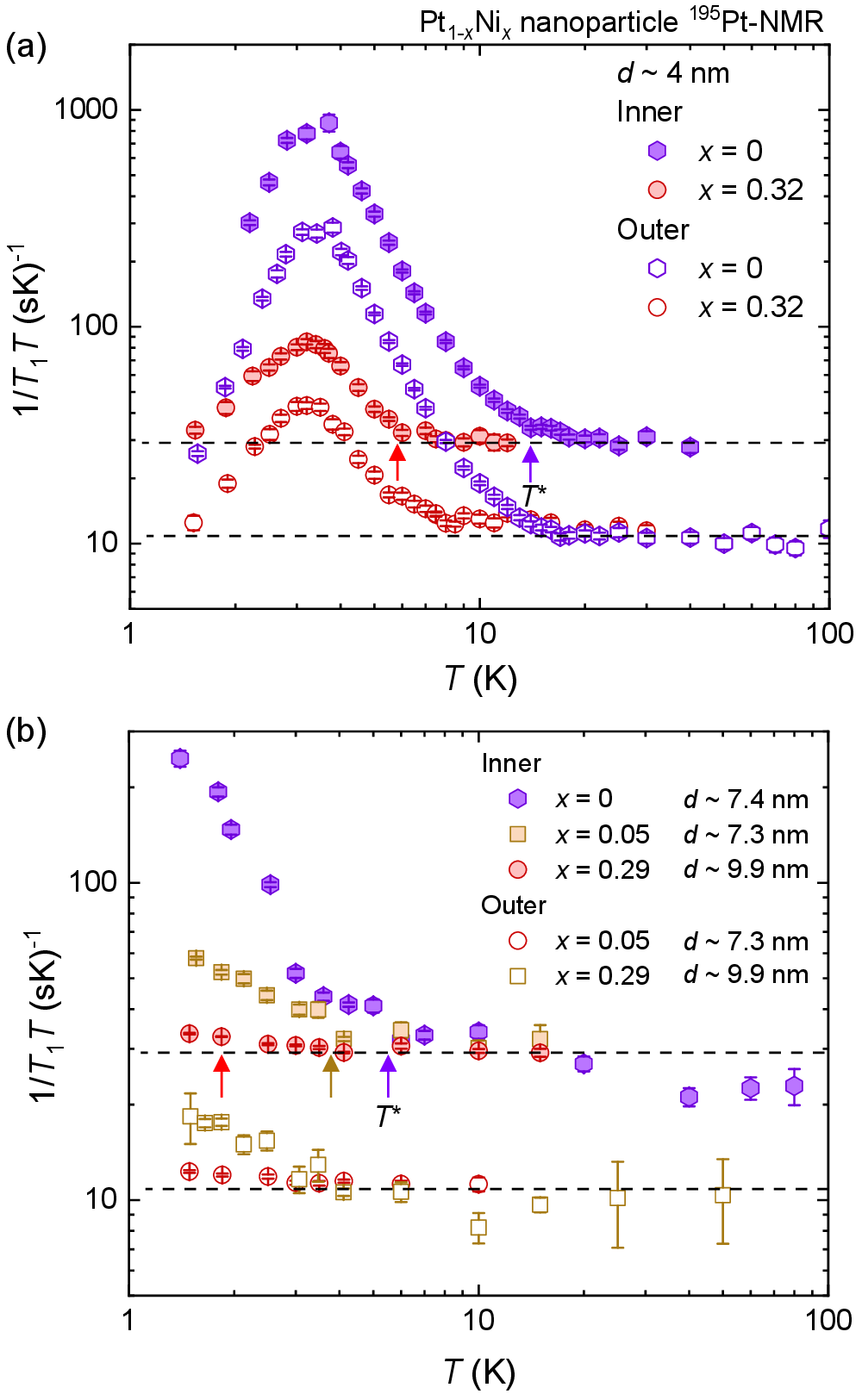}
\caption{
Temperature dependence of $1/T_1T$ in Pt$_{1-x}$Ni$_x$ nanoparticles at various $x$ for 4 nm (a) and 7-10~nm (b).
The dashed lines indicate a constant value at high temperatures.
The arrows indicate $T^*$, where $1/T_1T$ starts to increase.
For reference, the data for the monometallic Pt nanoparticles~\cite{T.Okuno_PRB_2020} are also presented.
}
\label{Fig.5}
\end{figure}

Figures~\ref{Fig.5}(a) and \ref{Fig.5}(b) present the temperature dependence of $1/T_1T$ measured at representative spectral positions corresponding to the inner and outer regions.
At $x = 0.3$ of 4~nm nanoparticles, the broad spectrum required the fitting with double Gaussian functions to estimate peak positions.
As in the Pt and Pt$_{1-x}$Pd$_x$ systems, $1/T_1T$ is constant at high temperatures, increases below $T^*$, and decreases at lower temperatures along with the maximum.
Although ferromagnetic fluctuations are enhanced in Pt$_{1-x}$Ni$_x$ nanoparticles, we attribute the low-temperature $1/T_1T$ increase to the QSE, rather than critical magnetic behavior, for the following reasons:
(1) No magnetic transition was detected in bulk magnetization measurements.
(2) The temperature dependence of $1/T_1T$ is almost the same between inner and outer regions.
(3) Measurements were conducted under high external magnetic fields ($\sim 12$~T), which would suppress the critical ferromagnetic fluctuations\cite{T.Hattori_PRL_2012}.
The value of $T^*$ decreases with increasing Ni content for both 4~nm and 7-10~nm nanoparticles.
This trend is interpreted as a consequence of increased $D(E_{\mathrm{F}})$ due to Ni substitution, leading to a smaller $\delta_{\mathrm{Kubo}}$.

\subsection{Comparison of $T^*$ and $\delta_{\mathrm{Kubo}}$}

\begin{figure}[!tb]
\includegraphics[width=\linewidth,clip]{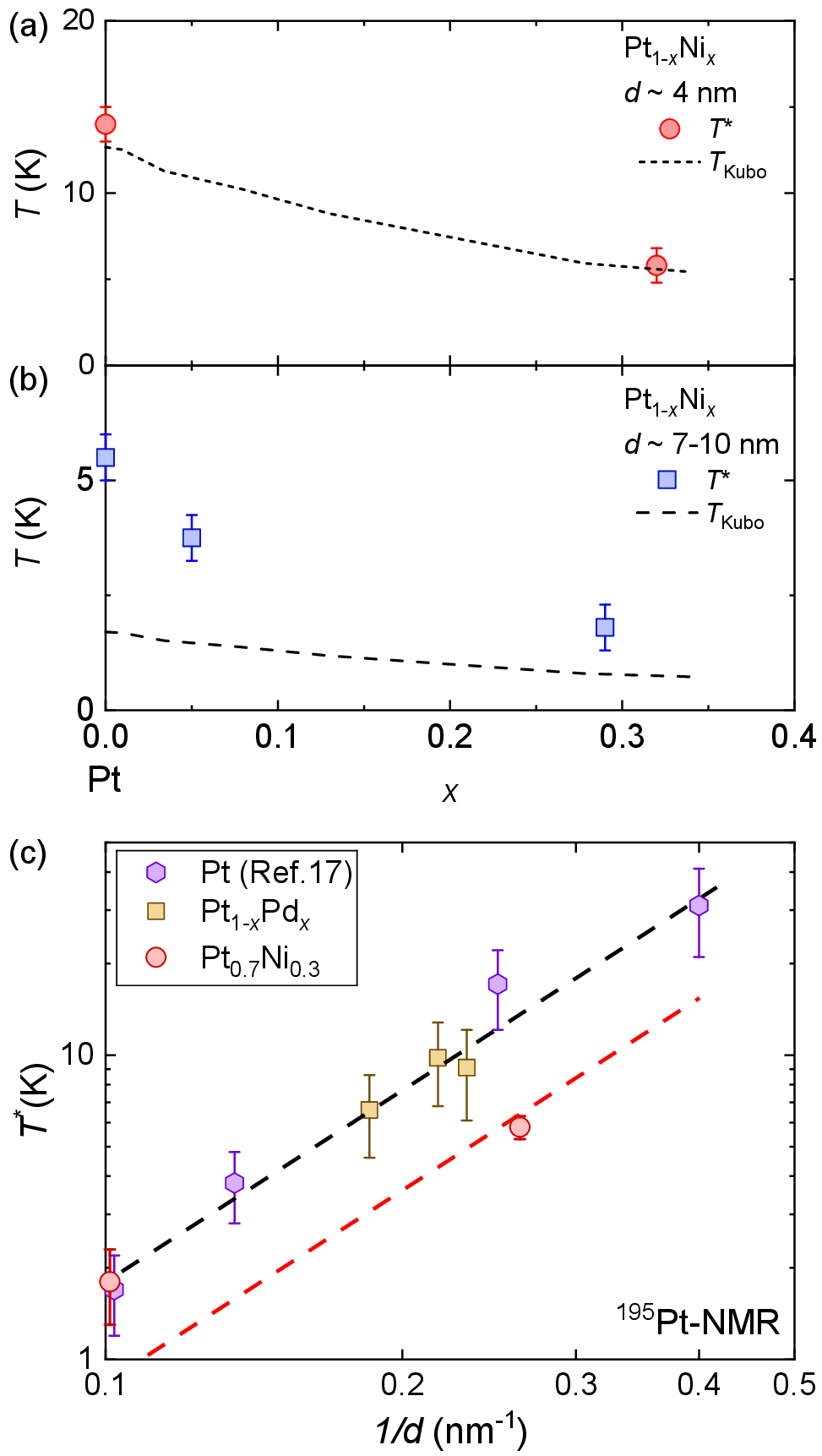}
\caption{
Ni composition $x$ dependence of $T^*$ in Pt$_{1-x}$Ni$_{x}$ nanoparticles for $d \sim 4$~nm (a) and 7-10~nm (b).
The dashed curves indicate a calculated $T_{\rm Kubo}$.
(c) $T^{*}$ vs. $1/d$ for various Pt-based nanoparticles.
The dashed lines are guides for the eyes.
For reference, the data for the monometallic Pt nanoparticles~\cite{T.Okuno_PRB_2020} are also presented.
}
\label{Fig.6}
\end{figure}

Figures~\ref{Fig.6}(a) and \ref{Fig.6}(b) compare the Ni concentration dependence of the observed $T^*$ and the estimated Kubo gap $\delta_{\mathrm{Kubo}}$ for 4~nm and 7-10~nm diameter nanoparticles, respectively.
Both quantities decrease with increasing $x$, in qualitative agreement with each other.
While $T^*$ is larger than $\delta_{\mathrm{Kubo}}/k_{\mathrm{B}}$, the consistent trend strongly supports the notion that $T^*$ originates from the QSE.
These observations stand in contrast to the behavior of Pt$_{1-x}$Cu$_x$ nanoparticles~\cite{S.Kitagawa_PRB_2024}, where the $s$-electron dominant Cu leads to the suppression of QSE signatures.
The present results highlight the critical role of $d$-electrons in the emergence of QSE in metallic nanoparticles.

\section{Discussion}

A key finding in this study is the systematic reduction in $T^*$ with increasing Ni concentration $x$.
This trend reflects the increase in $D(E_{\mathrm{F}})$ due to Ni substitution, which, according to the Kubo relation $\delta_{\mathrm{Kubo}} \propto 1/[N D(E_{\mathrm{F}})]$, results in a smaller energy gap.
As shown in Fig.~\ref{Fig.6}(c), this behavior is clearly illustrated by the linear relation between $T^*$ and the inverse particle diameter $1/d$.
For the Pt and Pt$_{1-x}$Pd$_x$ nanoparticles, data points with various diameters fall on a single straight line, consistent with the expectation from Kubo's theory.
This indicates that the electron states are similar between Pt and Pd atoms. 
In contrast, Pt$_{0.7}$Ni$_{0.3}$ nanoparticles exhibit a similar linear trend with a slightly lower $T^*$ at a given $1/d$, indicating a deviation from the universal line due to the enhanced $D(E_{\mathrm{F}})$ caused by Ni substitution.
This observation contrasts sharply with the previously reported behavior in Pt$_{1-x}$Cu$_x$ nanoparticles, where the substitution of Cu, whose conduction electrons are primarily of $s$-character, suppresses the low-temperature anomaly in $1/T_1$ and leads to a breakdown of the correlation with $\delta_{\mathrm{Kubo}}$~\cite{S.Kitagawa_PRB_2024}.
The comparison among Pt$_{1-x}$Cu$_x$, Pt$_{1-x}$Pd$_x$, and Pt$_{1-x}$Ni$_x$ nanoparticles clearly demonstrates that $d$-electron participation is essential for QSE to manifest.
These findings point toward a broader implication: the QSE in metallic nanoparticles is not only a geometric or size-dependent phenomenon but is also sensitive to the electronic structure and correlation effects of the constituent elements.
Particularly, in nanoparticles formed by metallic $d$-electron atoms, the presence of $d$-electrons near the Fermi level is a necessary condition for the discrete level structure to significantly influence low-temperature spin dynamics.
Such insights may have ramifications in the design of nanoscale quantum devices where fine control over quantum level spacing and electron correlations is required\cite{D.C.Ralph_PRL_1995}.

Recent studies further demonstrate that finite-size electronic states in metallic nanoparticles cannot always be characterized by a single mean level spacing. 
Theoretical calculations have shown that discrete electronic levels can produce pronounced magnetic field- and temperature-dependent structures in the magnetic susceptibility and that their observability is sensitive to thermal smearing and level broadening\cite{M.R-Llordes_PRB_2021}. 
Studies of atomically precise Au nanoclusters have also shown that the crossover between nonmetallic and metallic electronic structures depends not only on size but also on the experimental criterion used to define metallicity\cite{M.Zhou_ACSNano_2021}. 
Furthermore, spin-polarized calculations for Pt–Ni nanoalloys indicate that composition, morphology, and chemical ordering are closely coupled to their magnetic properties\cite{J.E.M.Cardona_JPCC_2023}.
These recent developments support the need to reconsider earlier QSE results obtained within a nearly-free-electron picture.
In $d$-electron nanoparticles, the strongly energy-dependent $D(E_{\mathrm{F}})$, electron correlations, spin-orbit coupling, surface coordination, and compositional inhomogeneity may renormalize or broaden the discrete electronic levels.
The present results show that the average scale $T^*$ remains related to $[ND(E_{\mathrm{F}})]^{-1}$, while the detailed low-temperature response cannot necessarily be determined by particle size alone.

\section{Conclusion}

We have investigated the QSE in the Pt$_{1-x}$Pd$_x$ and Pt$_{1-x}$Ni$_x$ nanoparticles using $^{195}$Pt NMR measurements.
The temperature dependence of $1/T_1T$ in the Pt$_{1-x}$Pd$_x$ nanoparticles exhibits a characteristic increase below $T^*$ followed by a maximum. 
This behavior is understood in terms of differences in particle size, as observed in Pt nanoparticles.
This indicates that the electronic states of the Pd atoms are almost the same as those of Pt atoms in the Pt$_{1-x}$Pd$_x$ nanoparticles.
In contrast, the temperature and $x$ dependences of $1/T_1T$ in the Pt$_{1-x}$Ni$_x$ nanoparticles show that $T^*$ systematically decreases with increasing Ni content, consistent with the Kubo gap $\delta_{\mathrm{Kubo}}$, although the maximum temperature of $1/T_1T$ is independent of $x$.
This behavior is in good agreement with the reduction in the Kubo gap $\delta_{\mathrm{Kubo}}$ expected from the enhanced $D(E_{\mathrm{F}})$ due to $d$-electron-rich Ni substitution, and the peak magnitude of $1/T_1T$ is governed primarily by particle size.
In contrast to the Pt$_{1-x}$Cu$_x$ nanoparticles, where the QSE is suppressed by $s$-electron dominant Cu, Pt$_{1-x}$Ni$_x$ nanoparticles retain the clear QSE signatures.
Furthermore, the modified Korringa analysis reveals enhanced ferromagnetic correlations at the higher Ni concentrations, which originate from the higher local density of states and further affect the QSE.
Our findings highlight the critical role of $d$-electrons in realizing QSE in the nanoparticles formed by metallic $d$-electron atoms.
This study suggests that quantum size effects in metallic nanoparticles should be reconsidered from the perspective of electron correlation and electronic structure.
Future investigations into other metallic $d$-electron nanoparticle systems may yield deeper insight into the interplay between QSE and many-body effects.

\section*{acknowledgments}
This work was supported by Grants-in-Aid for Scientific Research (KAKENHI Grant No. JP20KK0061, No. JP20H00130, No. JP21K18600, No. JP22H04933, No. JP22H01168, No. JP23H01124, No. JP23K22439, No. JP23K25821, and No. JP25H00609) from the Japan Society for the Promotion of Science, by research support funding from the Kyoto University Foundation, by ISHIZUE 2024 of Kyoto University Research Development Program, by Murata Science and Education Foundation, and by the JGC-S Scholarship Foundation.
Liquid helium is supplied by the Low Temperature and Materials Sciences Division, Agency for Health, Safety and Environment, Kyoto University.


\begin{thebibliography}{28}%
\makeatletter
\providecommand \@ifxundefined [1]{%
 \@ifx{#1\undefined}
}%
\providecommand \@ifnum [1]{%
 \ifnum #1\expandafter \@firstoftwo
 \else \expandafter \@secondoftwo
 \fi
}%
\providecommand \@ifx [1]{%
 \ifx #1\expandafter \@firstoftwo
 \else \expandafter \@secondoftwo
 \fi
}%
\providecommand \natexlab [1]{#1}%
\providecommand \enquote  [1]{``#1''}%
\providecommand \bibnamefont  [1]{#1}%
\providecommand \bibfnamefont [1]{#1}%
\providecommand \citenamefont [1]{#1}%
\providecommand \href@noop [0]{\@secondoftwo}%
\providecommand \href [0]{\begingroup \@sanitize@url \@href}%
\providecommand \@href[1]{\@@startlink{#1}\@@href}%
\providecommand \@@href[1]{\endgroup#1\@@endlink}%
\providecommand \@sanitize@url [0]{\catcode `\\12\catcode `\$12\catcode `\&12\catcode `\#12\catcode `\^12\catcode `\_12\catcode `\%12\relax}%
\providecommand \@@startlink[1]{}%
\providecommand \@@endlink[0]{}%
\providecommand \url  [0]{\begingroup\@sanitize@url \@url }%
\providecommand \@url [1]{\endgroup\@href {#1}{\urlprefix }}%
\providecommand \urlprefix  [0]{URL }%
\providecommand \Eprint [0]{\href }%
\providecommand \doibase [0]{https://doi.org/}%
\providecommand \selectlanguage [0]{\@gobble}%
\providecommand \bibinfo  [0]{\@secondoftwo}%
\providecommand \bibfield  [0]{\@secondoftwo}%
\providecommand \translation [1]{[#1]}%
\providecommand \BibitemOpen [0]{}%
\providecommand \bibitemStop [0]{}%
\providecommand \bibitemNoStop [0]{.\EOS\space}%
\providecommand \EOS [0]{\spacefactor3000\relax}%
\providecommand \BibitemShut  [1]{\csname bibitem#1\endcsname}%
\let\auto@bib@innerbib\@empty
\bibitem [{\citenamefont {Murray}\ \emph {et~al.}(2000)\citenamefont {Murray}, \citenamefont {Kagan},\ and\ \citenamefont {Bawendi}}]{C.B.Murray_ARMS_2000}%
  \BibitemOpen
  \bibfield  {author} {\bibinfo {author} {\bibfnamefont {C.~B.}\ \bibnamefont {Murray}}, \bibinfo {author} {\bibfnamefont {C.~R.}\ \bibnamefont {Kagan}},\ and\ \bibinfo {author} {\bibfnamefont {M.~G.}\ \bibnamefont {Bawendi}},\ }\bibfield  {title} {\bibinfo {title} {Synthesis and characterization of monodisperse nanocrystals and close-packed nanocrystal assemblies},\ }\href {https://doi.org/10.1146/annurev.matsci.30.1.545} {\bibfield  {journal} {\bibinfo  {journal} {Annu. Rev. Mater. Sci.}\ }\textbf {\bibinfo {volume} {30}},\ \bibinfo {pages} {545} (\bibinfo {year} {2000})}\BibitemShut {NoStop}%
\bibitem [{\citenamefont {Roduner}(2006)}]{E.Roduner_CSR_2006}%
  \BibitemOpen
  \bibfield  {author} {\bibinfo {author} {\bibfnamefont {E.}~\bibnamefont {Roduner}},\ }\bibfield  {title} {\bibinfo {title} {Size matters: why nanomaterials are different},\ }\href {https://doi.org/10.1039/b502142c} {\bibfield  {journal} {\bibinfo  {journal} {Chem. Soc. Rev.}\ }\textbf {\bibinfo {volume} {35}},\ \bibinfo {pages} {583} (\bibinfo {year} {2006})}\BibitemShut {NoStop}%
\bibitem [{\citenamefont {Ekimov}\ and\ \citenamefont {Onushchenko}(1981)}]{A.I.Ekimov_JETPL_1981}%
  \BibitemOpen
  \bibfield  {author} {\bibinfo {author} {\bibfnamefont {A.~I.}\ \bibnamefont {Ekimov}}\ and\ \bibinfo {author} {\bibfnamefont {A.~A.}\ \bibnamefont {Onushchenko}},\ }\bibfield  {title} {\bibinfo {title} {Quantum size effect in three-dimensional microscopic semiconductor crystals},\ }\href@noop {} {\bibfield  {journal} {\bibinfo  {journal} {JETP Letters}\ }\textbf {\bibinfo {volume} {34}},\ \bibinfo {pages} {345} (\bibinfo {year} {1981})}\BibitemShut {NoStop}%
\bibitem [{\citenamefont {Rossetti}\ \emph {et~al.}(1983)\citenamefont {Rossetti}, \citenamefont {Nakahara},\ and\ \citenamefont {Brus}}]{Rossetti1983}%
  \BibitemOpen
  \bibfield  {author} {\bibinfo {author} {\bibfnamefont {R.}~\bibnamefont {Rossetti}}, \bibinfo {author} {\bibfnamefont {S.}~\bibnamefont {Nakahara}},\ and\ \bibinfo {author} {\bibfnamefont {L.~E.}\ \bibnamefont {Brus}},\ }\bibfield  {title} {\bibinfo {title} {Quantum size effects in the redox potentials, resonance raman spectra, and electronic spectra of {CdS} crystallites in aqueous solution},\ }\href {https://doi.org/10.1063/1.445834} {\bibfield  {journal} {\bibinfo  {journal} {J. Chem. Phys.}\ }\textbf {\bibinfo {volume} {79}},\ \bibinfo {pages} {1086} (\bibinfo {year} {1983})}\BibitemShut {NoStop}%
\bibitem [{\citenamefont {Perenboom}\ \emph {et~al.}(1981)\citenamefont {Perenboom}, \citenamefont {Wyder},\ and\ \citenamefont {Meier}}]{J.A.A.J.Perenboom_PhysRep_1981}%
  \BibitemOpen
  \bibfield  {author} {\bibinfo {author} {\bibfnamefont {J.}~\bibnamefont {Perenboom}}, \bibinfo {author} {\bibfnamefont {P.}~\bibnamefont {Wyder}},\ and\ \bibinfo {author} {\bibfnamefont {F.}~\bibnamefont {Meier}},\ }\bibfield  {title} {\bibinfo {title} {Electronic properties of small metallic particles},\ }\href {https://doi.org/10.1016/0370-1573(81)90194-0} {\bibfield  {journal} {\bibinfo  {journal} {Phys. Rep.}\ }\textbf {\bibinfo {volume} {78}},\ \bibinfo {pages} {173} (\bibinfo {year} {1981})}\BibitemShut {NoStop}%
\bibitem [{\citenamefont {Kubo}(1962)}]{R.Kubo_JPSJ_1962}%
  \BibitemOpen
  \bibfield  {author} {\bibinfo {author} {\bibfnamefont {R.}~\bibnamefont {Kubo}},\ }\bibfield  {title} {\bibinfo {title} {Electronic properties of metallic fine particles. i.},\ }\href {https://doi.org/10.1143/JPSJ.17.975} {\bibfield  {journal} {\bibinfo  {journal} {J. Phys. Soc. Jpn.}\ }\textbf {\bibinfo {volume} {17}},\ \bibinfo {pages} {975} (\bibinfo {year} {1962})}\BibitemShut {NoStop}%
\bibitem [{\citenamefont {Gor'kov}\ and\ \citenamefont {Eliashberg}(1965)}]{L.P.Gorkov_JETP_1965}%
  \BibitemOpen
  \bibfield  {author} {\bibinfo {author} {\bibfnamefont {L.~P.}\ \bibnamefont {Gor'kov}}\ and\ \bibinfo {author} {\bibfnamefont {G.~M.}\ \bibnamefont {Eliashberg}},\ }\bibfield  {title} {\bibinfo {title} {Minute metallic particles in an electromagnetic field},\ }\href@noop {} {\bibfield  {journal} {\bibinfo  {journal} {Sov. Phys. JETP}\ }\textbf {\bibinfo {volume} {21}},\ \bibinfo {pages} {940} (\bibinfo {year} {1965})}\BibitemShut {NoStop}%
\bibitem [{\citenamefont {Halperin}(1986)}]{W.P.Halperin_RMP_1986}%
  \BibitemOpen
  \bibfield  {author} {\bibinfo {author} {\bibfnamefont {W.~P.}\ \bibnamefont {Halperin}},\ }\bibfield  {title} {\bibinfo {title} {Quantum size effects in metal particles},\ }\href {https://doi.org/10.1103/RevModPhys.58.533} {\bibfield  {journal} {\bibinfo  {journal} {Rev. Mod. Phys.}\ }\textbf {\bibinfo {volume} {58}},\ \bibinfo {pages} {533} (\bibinfo {year} {1986})}\BibitemShut {NoStop}%
\bibitem [{\citenamefont {Alivisatos}(1996)}]{A.P.Alivisatos_science_1996}%
  \BibitemOpen
  \bibfield  {author} {\bibinfo {author} {\bibfnamefont {A.~P.}\ \bibnamefont {Alivisatos}},\ }\bibfield  {title} {\bibinfo {title} {Semiconductor clusters, nanocrystals, and quantum dots},\ }\href {https://doi.org/10.1126/science.271.5251.933} {\bibfield  {journal} {\bibinfo  {journal} {Science}\ }\textbf {\bibinfo {volume} {271}},\ \bibinfo {pages} {933} (\bibinfo {year} {1996})}\BibitemShut {NoStop}%
\bibitem [{\citenamefont {Seh}\ \emph {et~al.}(2017)\citenamefont {Seh}, \citenamefont {Kibsgaard}, \citenamefont {Dickens}, \citenamefont {Chorkendorff}, \citenamefont {N{\o}rskov},\ and\ \citenamefont {Jaramillo}}]{Seh2017}%
  \BibitemOpen
  \bibfield  {author} {\bibinfo {author} {\bibfnamefont {Z.~W.}\ \bibnamefont {Seh}}, \bibinfo {author} {\bibfnamefont {J.}~\bibnamefont {Kibsgaard}}, \bibinfo {author} {\bibfnamefont {C.~F.}\ \bibnamefont {Dickens}}, \bibinfo {author} {\bibfnamefont {I.}~\bibnamefont {Chorkendorff}}, \bibinfo {author} {\bibfnamefont {J.~K.}\ \bibnamefont {N{\o}rskov}},\ and\ \bibinfo {author} {\bibfnamefont {T.~F.}\ \bibnamefont {Jaramillo}},\ }\bibfield  {title} {\bibinfo {title} {Combining theory and experiment in electrocatalysis: Insights into materials design},\ }\href@noop {} {\bibfield  {journal} {\bibinfo  {journal} {Science}\ }\textbf {\bibinfo {volume} {355}},\ \bibinfo {pages} {146} (\bibinfo {year} {2017})}\BibitemShut {NoStop}%
\bibitem [{\citenamefont {Harish}\ \emph {et~al.}(2022)\citenamefont {Harish}, \citenamefont {Tewari}, \citenamefont {Gaur}, \citenamefont {Yadav}, \citenamefont {Swaroop}, \citenamefont {Bechelany},\ and\ \citenamefont {Barhoum}}]{V.Harish_nano_2022}%
  \BibitemOpen
  \bibfield  {author} {\bibinfo {author} {\bibfnamefont {V.}~\bibnamefont {Harish}}, \bibinfo {author} {\bibfnamefont {D.}~\bibnamefont {Tewari}}, \bibinfo {author} {\bibfnamefont {M.}~\bibnamefont {Gaur}}, \bibinfo {author} {\bibfnamefont {A.~B.}\ \bibnamefont {Yadav}}, \bibinfo {author} {\bibfnamefont {S.}~\bibnamefont {Swaroop}}, \bibinfo {author} {\bibfnamefont {M.}~\bibnamefont {Bechelany}},\ and\ \bibinfo {author} {\bibfnamefont {A.}~\bibnamefont {Barhoum}},\ }\bibfield  {title} {\bibinfo {title} {Review on nanoparticles and nanostructured materials: Bioimaging, biosensing, drug delivery, tissue engineering, antimicrobial, and agro-food applications},\ }\href {https://doi.org/10.3390/nano12030457} {\bibfield  {journal} {\bibinfo  {journal} {Nanomaterials}\ }\textbf {\bibinfo {volume} {12}},\ \bibinfo {pages} {457} (\bibinfo {year} {2022})}\BibitemShut {NoStop}%
\bibitem [{\citenamefont {Antoine}\ \emph {et~al.}(2023)\citenamefont {Antoine}, \citenamefont {Broyer},\ and\ \citenamefont {Dugourd}}]{Antoine2023}%
  \BibitemOpen
  \bibfield  {author} {\bibinfo {author} {\bibfnamefont {R.}~\bibnamefont {Antoine}}, \bibinfo {author} {\bibfnamefont {M.}~\bibnamefont {Broyer}},\ and\ \bibinfo {author} {\bibfnamefont {P.}~\bibnamefont {Dugourd}},\ }\bibfield  {title} {\bibinfo {title} {Metal nanoclusters: from fundamental aspects to electronic properties and optical applications},\ }\href {https://doi.org/10.1080/14686996.2023.2222546} {\bibfield  {journal} {\bibinfo  {journal} {Sci. Technol. Adv. Mater.}\ }\textbf {\bibinfo {volume} {24}},\ \bibinfo {pages} {1} (\bibinfo {year} {2023})}\BibitemShut {NoStop}%
\bibitem [{\citenamefont {Volokitin}\ \emph {et~al.}(1996)\citenamefont {Volokitin}, \citenamefont {Sinzig}, \citenamefont {de~Jongh}, \citenamefont {Schmid}, \citenamefont {Vargaftik},\ and\ \citenamefont {Moiseevi}}]{Y.Volokitin_Nature_1996}%
  \BibitemOpen
  \bibfield  {author} {\bibinfo {author} {\bibfnamefont {Y.}~\bibnamefont {Volokitin}}, \bibinfo {author} {\bibfnamefont {J.}~\bibnamefont {Sinzig}}, \bibinfo {author} {\bibfnamefont {L.~J.}\ \bibnamefont {de~Jongh}}, \bibinfo {author} {\bibfnamefont {G.}~\bibnamefont {Schmid}}, \bibinfo {author} {\bibfnamefont {M.~N.}\ \bibnamefont {Vargaftik}},\ and\ \bibinfo {author} {\bibfnamefont {I.~I.}\ \bibnamefont {Moiseevi}},\ }\bibfield  {title} {\bibinfo {title} {Quantum-size effects in the thermodynamic properties of metallic nanoparticles},\ }\href {https://doi.org/10.1038/384621a0} {\bibfield  {journal} {\bibinfo  {journal} {Nature}\ }\textbf {\bibinfo {volume} {384}},\ \bibinfo {pages} {621} (\bibinfo {year} {1996})}\BibitemShut {NoStop}%
\bibitem [{\citenamefont {Bucher}\ \emph {et~al.}(1989)\citenamefont {Bucher}, \citenamefont {Buttet}, \citenamefont {Klink},\ and\ \citenamefont {Graetzel}}]{Bucher1989}%
  \BibitemOpen
  \bibfield  {author} {\bibinfo {author} {\bibfnamefont {J.}~\bibnamefont {Bucher}}, \bibinfo {author} {\bibfnamefont {J.}~\bibnamefont {Buttet}}, \bibinfo {author} {\bibfnamefont {J.~V.~D.}\ \bibnamefont {Klink}},\ and\ \bibinfo {author} {\bibfnamefont {M.}~\bibnamefont {Graetzel}},\ }\bibfield  {title} {\bibinfo {title} {Electronic properties and local densities of states in clean and hydrogen covered pt particles},\ }\href {https://doi.org/10.1016/0039-6028(89)90175-1} {\bibfield  {journal} {\bibinfo  {journal} {Surf. Sci.}\ }\textbf {\bibinfo {volume} {214}},\ \bibinfo {pages} {347} (\bibinfo {year} {1989})}\BibitemShut {NoStop}%
\bibitem [{\citenamefont {Fujii}\ \emph {et~al.}(2022)\citenamefont {Fujii}, \citenamefont {Iwamoto}, \citenamefont {Nakai}, \citenamefont {Shiratsu}, \citenamefont {Yao}, \citenamefont {Ueda},\ and\ \citenamefont {Mito}}]{T.Fujii_PRB_2022}%
  \BibitemOpen
  \bibfield  {author} {\bibinfo {author} {\bibfnamefont {T.}~\bibnamefont {Fujii}}, \bibinfo {author} {\bibfnamefont {K.}~\bibnamefont {Iwamoto}}, \bibinfo {author} {\bibfnamefont {Y.}~\bibnamefont {Nakai}}, \bibinfo {author} {\bibfnamefont {T.}~\bibnamefont {Shiratsu}}, \bibinfo {author} {\bibfnamefont {H.}~\bibnamefont {Yao}}, \bibinfo {author} {\bibfnamefont {K.}~\bibnamefont {Ueda}},\ and\ \bibinfo {author} {\bibfnamefont {T.}~\bibnamefont {Mito}},\ }\bibfield  {title} {\bibinfo {title} {{NMR} evidence for energy gap opening in thiol-capped platinum nanoparticles},\ }\href {https://doi.org/10.1103/PhysRevB.105.L121401} {\bibfield  {journal} {\bibinfo  {journal} {Phys. Rev. B}\ }\textbf {\bibinfo {volume} {105}},\ \bibinfo {pages} {L121401} (\bibinfo {year} {2022})}\BibitemShut {NoStop}%
\bibitem [{\citenamefont {Rhodes}\ \emph {et~al.}(1982)\citenamefont {Rhodes}, \citenamefont {Wang}, \citenamefont {Stokes}, \citenamefont {Slichter},\ and\ \citenamefont {Sinfelt}}]{H.E.Rhodes_PRB_1982}%
  \BibitemOpen
  \bibfield  {author} {\bibinfo {author} {\bibfnamefont {H.~E.}\ \bibnamefont {Rhodes}}, \bibinfo {author} {\bibfnamefont {P.-K.}\ \bibnamefont {Wang}}, \bibinfo {author} {\bibfnamefont {H.~T.}\ \bibnamefont {Stokes}}, \bibinfo {author} {\bibfnamefont {C.~P.}\ \bibnamefont {Slichter}},\ and\ \bibinfo {author} {\bibfnamefont {J.~H.}\ \bibnamefont {Sinfelt}},\ }\bibfield  {title} {\bibinfo {title} {{NMR} of platinum catalysts. i. line shapes},\ }\href {https://doi.org/10.1103/PhysRevB.26.3559} {\bibfield  {journal} {\bibinfo  {journal} {Phys. Rev. B}\ }\textbf {\bibinfo {volume} {26}},\ \bibinfo {pages} {3559} (\bibinfo {year} {1982})}\BibitemShut {NoStop}%
\bibitem [{\citenamefont {Okuno}\ \emph {et~al.}(2020{\natexlab{a}})\citenamefont {Okuno}, \citenamefont {Manago}, \citenamefont {Kitagawa}, \citenamefont {Ishida}, \citenamefont {Kusada},\ and\ \citenamefont {Kitagawa}}]{T.Okuno_PRB_2020}%
  \BibitemOpen
  \bibfield  {author} {\bibinfo {author} {\bibfnamefont {T.}~\bibnamefont {Okuno}}, \bibinfo {author} {\bibfnamefont {M.}~\bibnamefont {Manago}}, \bibinfo {author} {\bibfnamefont {S.}~\bibnamefont {Kitagawa}}, \bibinfo {author} {\bibfnamefont {K.}~\bibnamefont {Ishida}}, \bibinfo {author} {\bibfnamefont {K.}~\bibnamefont {Kusada}},\ and\ \bibinfo {author} {\bibfnamefont {H.}~\bibnamefont {Kitagawa}},\ }\bibfield  {title} {\bibinfo {title} {Nmr-based gap behavior related to the quantum size effect},\ }\href {https://doi.org/10.1103/PhysRevB.101.121406} {\bibfield  {journal} {\bibinfo  {journal} {Phys. Rev. B}\ }\textbf {\bibinfo {volume} {101}},\ \bibinfo {pages} {121406} (\bibinfo {year} {2020}{\natexlab{a}})}\BibitemShut {NoStop}%
\bibitem [{\citenamefont {Okuno}\ \emph {et~al.}(2020{\natexlab{b}})\citenamefont {Okuno}, \citenamefont {Kinoshita}, \citenamefont {Matsuzaki}, \citenamefont {Kitagawa}, \citenamefont {Ishida}, \citenamefont {Hirata}, \citenamefont {Sasaki}, \citenamefont {Kusada},\ and\ \citenamefont {Kitagawa}}]{T.Okuno_JPSJ_2020}%
  \BibitemOpen
  \bibfield  {author} {\bibinfo {author} {\bibfnamefont {T.}~\bibnamefont {Okuno}}, \bibinfo {author} {\bibfnamefont {Y.}~\bibnamefont {Kinoshita}}, \bibinfo {author} {\bibfnamefont {S.}~\bibnamefont {Matsuzaki}}, \bibinfo {author} {\bibfnamefont {S.}~\bibnamefont {Kitagawa}}, \bibinfo {author} {\bibfnamefont {K.}~\bibnamefont {Ishida}}, \bibinfo {author} {\bibfnamefont {M.}~\bibnamefont {Hirata}}, \bibinfo {author} {\bibfnamefont {T.}~\bibnamefont {Sasaki}}, \bibinfo {author} {\bibfnamefont {K.}~\bibnamefont {Kusada}},\ and\ \bibinfo {author} {\bibfnamefont {H.}~\bibnamefont {Kitagawa}},\ }\bibfield  {title} {\bibinfo {title} {Magnetic-field dependence of novel gap behavior related to the quantum-size effect},\ }\href {https://doi.org/10.7566/JPSJ.89.095002} {\bibfield  {journal} {\bibinfo  {journal} {J. Phys. Soc. Jpn.}\ }\textbf {\bibinfo {volume} {89}},\ \bibinfo {pages} {095002} (\bibinfo {year} {2020}{\natexlab{b}})}\BibitemShut {NoStop}%
\bibitem [{\citenamefont {Hammer}\ and\ \citenamefont {Norskov}(1995)}]{B.Hammer_Nature_1995}%
  \BibitemOpen
  \bibfield  {author} {\bibinfo {author} {\bibfnamefont {B.}~\bibnamefont {Hammer}}\ and\ \bibinfo {author} {\bibfnamefont {J.~K.}\ \bibnamefont {Norskov}},\ }\bibfield  {title} {\bibinfo {title} {Why gold is the noblest of all the metals},\ }\href {https://doi.org/10.1038/376238a0} {\bibfield  {journal} {\bibinfo  {journal} {Nature}\ }\textbf {\bibinfo {volume} {376}},\ \bibinfo {pages} {238} (\bibinfo {year} {1995})}\BibitemShut {NoStop}%
\bibitem [{\citenamefont {Kitagawa}\ \emph {et~al.}(2024)\citenamefont {Kitagawa}, \citenamefont {Kinoshita}, \citenamefont {Ishida}, \citenamefont {Kusada},\ and\ \citenamefont {Kitagawa}}]{S.Kitagawa_PRB_2024}%
  \BibitemOpen
  \bibfield  {author} {\bibinfo {author} {\bibfnamefont {S.}~\bibnamefont {Kitagawa}}, \bibinfo {author} {\bibfnamefont {Y.}~\bibnamefont {Kinoshita}}, \bibinfo {author} {\bibfnamefont {K.}~\bibnamefont {Ishida}}, \bibinfo {author} {\bibfnamefont {K.}~\bibnamefont {Kusada}},\ and\ \bibinfo {author} {\bibfnamefont {H.}~\bibnamefont {Kitagawa}},\ }\bibfield  {title} {\bibinfo {title} {Breakdown of kubo relation in pt-cu nanoparticles},\ }\href {https://doi.org/10.1103/PhysRevB.109.L041408} {\bibfield  {journal} {\bibinfo  {journal} {Phys. Rev. B}\ }\textbf {\bibinfo {volume} {109}},\ \bibinfo {pages} {L041408} (\bibinfo {year} {2024})}\BibitemShut {NoStop}%
\bibitem [{\citenamefont {Inoue}\ and\ \citenamefont {Shimizu}(1977)}]{J.Inoue_JPSJ_1977}%
  \BibitemOpen
  \bibfield  {author} {\bibinfo {author} {\bibfnamefont {J.}~\bibnamefont {Inoue}}\ and\ \bibinfo {author} {\bibfnamefont {M.}~\bibnamefont {Shimizu}},\ }\bibfield  {title} {\bibinfo {title} {Magnetic properties of ni-pd, ni-pt and pd-pt alloys},\ }\href {https://doi.org/10.1143/JPSJ.42.1547} {\bibfield  {journal} {\bibinfo  {journal} {J. Phys. Soc. Jpn.}\ }\textbf {\bibinfo {volume} {42}},\ \bibinfo {pages} {1547} (\bibinfo {year} {1977})}\BibitemShut {NoStop}%
\bibitem [{\citenamefont {Batlle}\ and\ \citenamefont {Labarta}(2002)}]{B.XavierBatlle_JPhysD_2002}%
  \BibitemOpen
  \bibfield  {author} {\bibinfo {author} {\bibfnamefont {X.}~\bibnamefont {Batlle}}\ and\ \bibinfo {author} {\bibfnamefont {A.}~\bibnamefont {Labarta}},\ }\bibfield  {title} {\bibinfo {title} {Finite-size effects in fine particles: magnetic and transport properties},\ }\href {https://doi.org/10.1088/0022-3727/35/6/201} {\bibfield  {journal} {\bibinfo  {journal} {J. Phys. D: Appl. Phys.}\ }\textbf {\bibinfo {volume} {35}},\ \bibinfo {pages} {201} (\bibinfo {year} {2002})}\BibitemShut {NoStop}%
\bibitem [{\citenamefont {Harris}\ \emph {et~al.}(2001)\citenamefont {Harris}, \citenamefont {Becker}, \citenamefont {Cabral~de Menezes}, \citenamefont {Goodfellow},\ and\ \citenamefont {Granger}}]{R.K.Harris_2001}%
  \BibitemOpen
  \bibfield  {author} {\bibinfo {author} {\bibfnamefont {R.~K.}\ \bibnamefont {Harris}}, \bibinfo {author} {\bibfnamefont {E.~D.}\ \bibnamefont {Becker}}, \bibinfo {author} {\bibfnamefont {S.~M.}\ \bibnamefont {Cabral~de Menezes}}, \bibinfo {author} {\bibfnamefont {R.}~\bibnamefont {Goodfellow}},\ and\ \bibinfo {author} {\bibfnamefont {P.}~\bibnamefont {Granger}},\ }\bibfield  {title} {\bibinfo {title} {{NMR nomenclature. Nuclear spin properties and conventions for chemical shifts(IUPAC Recommendations 2001)}},\ }\href {https://doi.org/10.1351/pac200173111795} {\bibfield  {journal} {\bibinfo  {journal} {Pure Appl. Chem.}\ }\textbf {\bibinfo {volume} {73}},\ \bibinfo {pages} {1795} (\bibinfo {year} {2001})}\BibitemShut {NoStop}%
\bibitem [{\citenamefont {Stone}(2016)}]{N.J.Stone_Q_2016}%
  \BibitemOpen
  \bibfield  {author} {\bibinfo {author} {\bibfnamefont {N.~J.}\ \bibnamefont {Stone}},\ }\bibfield  {title} {\bibinfo {title} {Table of nuclear electric quadrupole moments},\ }\href {https://doi.org/10.1016/j.adt.2015.12.002} {\bibfield  {journal} {\bibinfo  {journal} {At. Data Nucl. Data Tables}\ }\textbf {\bibinfo {volume} {111--112}},\ \bibinfo {pages} {1} (\bibinfo {year} {2016})}\BibitemShut {NoStop}%
\bibitem [{\citenamefont {Carter}\ \emph {et~al.}(1977)\citenamefont {Carter}, \citenamefont {Bennet},\ and\ \citenamefont {Kahan}}]{Metallicshifts_1977}%
  \BibitemOpen
  \bibfield  {author} {\bibinfo {author} {\bibfnamefont {G.~C.}\ \bibnamefont {Carter}}, \bibinfo {author} {\bibfnamefont {L.~H.}\ \bibnamefont {Bonnet}},\ and\ \bibinfo {author} {\bibfnamefont {D.~J.}\ \bibnamefont {Kahan}},\ }\href@noop {} {\emph {\bibinfo {title} {Metallic Shifts in NMR}}}\ (\bibinfo  {publisher} {Pergamon Press, Oxford},\ \bibinfo {year} {1977})\BibitemShut {NoStop}%
\bibitem [{\citenamefont {Moriya}(1963)}]{T.Moriya_JPSJ_1963}%
  \BibitemOpen
  \bibfield  {author} {\bibinfo {author} {\bibfnamefont {T.}~\bibnamefont {Moriya}},\ }\bibfield  {title} {\bibinfo {title} {The {E}ffect of {E}lectron-{E}lectron {I}nteraction on the {N}uclear {S}pin {R}elaxation in {M}etals},\ }\href@noop {} {\bibfield  {journal} {\bibinfo  {journal} {J. Phys. Soc. Jpn.}\ }\textbf {\bibinfo {volume} {18}},\ \bibinfo {pages} {516} (\bibinfo {year} {1963})}\BibitemShut {NoStop}%
\bibitem [{\citenamefont {Hattori}\ \emph {et~al.}(2012)\citenamefont {Hattori}, \citenamefont {Ihara}, \citenamefont {Nakai}, \citenamefont {Ishida}, \citenamefont {Tada}, \citenamefont {Fujimoto}, \citenamefont {Kawakami}, \citenamefont {Osaki}, \citenamefont {Deguchi}, \citenamefont {Sato},\ and\ \citenamefont {Satoh}}]{T.Hattori_PRL_2012}%
  \BibitemOpen
  \bibfield  {author} {\bibinfo {author} {\bibfnamefont {T.}~\bibnamefont {Hattori}}, \bibinfo {author} {\bibfnamefont {Y.}~\bibnamefont {Ihara}}, \bibinfo {author} {\bibfnamefont {Y.}~\bibnamefont {Nakai}}, \bibinfo {author} {\bibfnamefont {K.}~\bibnamefont {Ishida}}, \bibinfo {author} {\bibfnamefont {Y.}~\bibnamefont {Tada}}, \bibinfo {author} {\bibfnamefont {S.}~\bibnamefont {Fujimoto}}, \bibinfo {author} {\bibfnamefont {N.}~\bibnamefont {Kawakami}}, \bibinfo {author} {\bibfnamefont {E.}~\bibnamefont {Osaki}}, \bibinfo {author} {\bibfnamefont {K.}~\bibnamefont {Deguchi}}, \bibinfo {author} {\bibfnamefont {N.~K.}\ \bibnamefont {Sato}},\ and\ \bibinfo {author} {\bibfnamefont {I.}~\bibnamefont {Satoh}},\ }\bibfield  {title} {\bibinfo {title} {Superconductivity induced by longitudinal ferromagnetic fluctuations in ucoge},\ }\href {https://doi.org/10.1103/PhysRevLett.108.066403} {\bibfield  {journal} {\bibinfo  {journal} {Phys. Rev. Lett.}\ }\textbf {\bibinfo {volume} {108}},\ \bibinfo {pages} {066403}
  (\bibinfo {year} {2012})}\BibitemShut {NoStop}%
\bibitem [{\citenamefont {Ralph}\ \emph {et~al.}(1995)\citenamefont {Ralph}, \citenamefont {Black},\ and\ \citenamefont {Tinkham}}]{D.C.Ralph_PRL_1995}%
  \BibitemOpen
  \bibfield  {author} {\bibinfo {author} {\bibfnamefont {D.~C.}\ \bibnamefont {Ralph}}, \bibinfo {author} {\bibfnamefont {C.~T.}\ \bibnamefont {Black}},\ and\ \bibinfo {author} {\bibfnamefont {M.}~\bibnamefont {Tinkham}},\ }\bibfield  {title} {\bibinfo {title} {Spectroscopic measurements of discrete electronic states in single metal particles},\ }\href {https://doi.org/10.1103/PhysRevLett.74.3241} {\bibfield  {journal} {\bibinfo  {journal} {Phys. Rev. Lett.}\ }\textbf {\bibinfo {volume} {74}},\ \bibinfo {pages} {3241} (\bibinfo {year} {1995})}\BibitemShut {NoStop}%
\bibitem [{\citenamefont {Roda-Llordes}\ \emph {et~al.}(2021)\citenamefont
  {Roda-Llordes}, \citenamefont {Gonzalez-Ballestero}, \citenamefont {L\'opez},
  \citenamefont {Mart\'{\i}nez-P\'erez}, \citenamefont {Luis},\ and\
  \citenamefont {Romero-Isart}}]{M.R-Llordes_PRB_2021}%
  \BibitemOpen
  \bibfield  {author} {\bibinfo {author} {\bibfnamefont {M.}~\bibnamefont
  {Roda-Llordes}}, \bibinfo {author} {\bibfnamefont {C.}~\bibnamefont
  {Gonzalez-Ballestero}}, \bibinfo {author} {\bibfnamefont {A.~E.~R.}\
  \bibnamefont {L\'opez}}, \bibinfo {author} {\bibfnamefont {M.~J.}\
  \bibnamefont {Mart\'{\i}nez-P\'erez}}, \bibinfo {author} {\bibfnamefont
  {F.}~\bibnamefont {Luis}},\ and\ \bibinfo {author} {\bibfnamefont
  {O.}~\bibnamefont {Romero-Isart}},\ }\bibfield  {title} {\bibinfo {title}
  {Quantum size effects in the magnetic susceptibility of a metallic
  nanoparticle},\ }\href {https://doi.org/10.1103/PhysRevB.104.L100407}
  {\bibfield  {journal} {\bibinfo  {journal} {Phys. Rev. B}\ }\textbf {\bibinfo
  {volume} {104}},\ \bibinfo {pages} {L100407} (\bibinfo {year}
  {2021})}\BibitemShut {NoStop}%
\bibitem [{\citenamefont {Zhou}\ \emph {et~al.}(2021)\citenamefont {Zhou},
  \citenamefont {Du}, \citenamefont {Wang},\ and\ \citenamefont
  {Jin}}]{M.Zhou_ACSNano_2021}%
  \BibitemOpen
  \bibfield  {author} {\bibinfo {author} {\bibfnamefont {M.}~\bibnamefont
  {Zhou}}, \bibinfo {author} {\bibfnamefont {X.}~\bibnamefont {Du}}, \bibinfo
  {author} {\bibfnamefont {H.}~\bibnamefont {Wang}},\ and\ \bibinfo {author}
  {\bibfnamefont {R.}~\bibnamefont {Jin}},\ }\bibfield  {title} {\bibinfo
  {title} {The critical number of gold atoms for a metallic state nanocluster:
  Resolving a decades-long question},\ }\href
  {https://doi.org/10.1021/acsnano.1c04705} {\bibfield  {journal} {\bibinfo
  {journal} {ACS Nano}\ }\textbf {\bibinfo {volume} {15}},\ \bibinfo {pages}
  {13980} (\bibinfo {year} {2021})}\BibitemShut {NoStop}%
\bibitem [{\citenamefont {Cardona}\ \emph {et~al.}(2023)\citenamefont
  {Cardona}, \citenamefont {Salichon}, \citenamefont {Tarrat}, \citenamefont
  {Gaudry},\ and\ \citenamefont {Loffreda}}]{J.E.M.Cardona_JPCC_2023}%
  \BibitemOpen
  \bibfield  {author} {\bibinfo {author} {\bibfnamefont {J.~E.~M.}\
  \bibnamefont {Cardona}}, \bibinfo {author} {\bibfnamefont {A.}~\bibnamefont
  {Salichon}}, \bibinfo {author} {\bibfnamefont {N.}~\bibnamefont {Tarrat}},
  \bibinfo {author} {\bibfnamefont {E.}~\bibnamefont {Gaudry}},\ and\ \bibinfo
  {author} {\bibfnamefont {D.}~\bibnamefont {Loffreda}},\ }\bibfield  {title}
  {\bibinfo {title} {Structural, ordering, and magnetic properties of ptni
  nanoalloys explored by density functional theory and stability descriptors},\
  }\href {https://doi.org/10.1021/acs.jpcc.3c03541} {\bibfield  {journal}
  {\bibinfo  {journal} {J. Phys. Chem. C}\ }\textbf {\bibinfo {volume} {127}},\
  \bibinfo {pages} {18043} (\bibinfo {year} {2023})}\BibitemShut {NoStop}%
\end{thebibliography}
\end{document}